\documentclass[a4paper,onecolumn,11pt, noarxiv, allowtoday]{quantumarticle}
\pdfoutput=1
\usepackage[hidelinks]{hyperref}

\usepackage{graphicx}
\usepackage{color}
\usepackage{amsmath}
\usepackage{amsfonts}
\usepackage{amssymb}
\usepackage{float}

\begin{document}
\title{Simulating quantum circuit expectation values by Clifford perturbation theory}
\author{Tomislav Begu\v{s}i\'{c}}\thanks{These authors contributed equally to this work.}
\author{Kasra Hejazi}\thanks{These authors contributed equally to this work.}
\author{Garnet Kin-Lic Chan}
\email{gkc1000@gmail.com}
\affiliation{Division of Chemistry and Chemical Engineering, California Institute of Technology, Pasadena, California 91125, USA}
\date{\today}

\begin{abstract}
The classical simulation of quantum circuits is of central importance for benchmarking near-term quantum devices. The fact that gates belonging to the Clifford group can be simulated efficiently on classical computers has motivated a range of methods that scale exponentially only in the number of non-Clifford gates. Here, we consider the expectation value problem for circuits composed of Clifford gates and non-Clifford Pauli rotations, and introduce a heuristic perturbative approach based on the truncation of the exponentially growing sum of Pauli terms in the Heisenberg picture. Numerical results are shown on a Quantum Approximate Optimization Algorithm (QAOA) benchmark for the E3LIN2 problem
and we also demonstrate how this method can be used to quantify coherent and incoherent errors of local observables in Clifford circuits.
Our results indicate that this systematically improvable perturbative method offers a viable alternative to exact methods for approximating expectation values of large near-Clifford circuits.
\end{abstract}

\maketitle

\graphicspath{{./figures}}

\section{Introduction\label{sec:intro}}

Validating near-term quantum devices with classical simulations is paramount for their future development \cite{Cross_Gambetta:2019}. In general, the classical simulation of quantum circuits is hard and thus limited to a small number of qubits or gates. One exception is Clifford circuits, which can be simulated efficiently using the stabilizer formalism~\cite{Gottesman:1998,Aaronson_Gottesman:2004,VanDenNest:2010}. 
From this starting point, circuit simulation methods have been developed
which scale exponentially only with the number of non-Clifford gates~\cite{Bravyi_Gosset:2016,Bennink_Pooser:2017,Qassim_Emerson:2019,Bravyi_Howard:2019,Huang_Love:2021,Kissinger_Wetering:2022}.
Currently, the low-rank stabilizer method proposed by Bravyi et al. \cite{Bravyi_Howard:2019} is one of the best-scaling methods for sampling the output distribution of a quantum circuit with a small number of non-Clifford gates.


In this work, we are concerned not with sampling the full output distribution of a quantum circuit, but only with evaluating the expectation value (mean value) of Pauli operators. 
Quantum mean values~\cite{Bravyi_Movassagh:2021} are not only a central aspect of the quantum output, but also form the cost function
 in variational quantum algorithms \cite{Cerezo_Coles:2021}, such as the variational quantum eigensolver (VQE) \cite{Peruzzo_OBrien:2014} or QAOA \cite{Farhi_Gutmann:2014, Farhi_Harrow:2016}.
Here, we introduce a technique to obtain expectation values of quantum circuits based on a
Clifford-based perturbation method. The method is most efficient when the gates are close to Clifford. Specifically, in the proposed heuristic inspired by time-dependent perturbation theory, the general Heisenberg evolution of a Pauli observable is computed as a polynomial in non-Clifford parameters, and the polynomial is truncated to control the cost. We use numerical examples to show that accurate results can be obtained even with low orders of perturbation, and we demonstrate on a QAOA benchmark that the method is faster than computing expectation values by sampling methods
by orders of magnitude. Finally, we show how the method can be used to model coherent and incoherent noise in Clifford circuits, which is relevant to benchmarking stabilizer error-correcting codes with many qubits.


\section{Method\label{sec:method}}

We consider the expectation value
\begin{equation}
\langle O \rangle = \langle 0^{\otimes n}| C_N^{\dag}  U_N^{\dag} \cdots C_1^{\dag} U_1^{\dag} O U_1 C_1 \cdots U_N C_N |0^{\otimes n} \rangle \label{eq:exp_val_m}
\end{equation}
of a Pauli operator $O$ evolved under a set of general unitary operators (quantum gates) $U_i$ and Clifford gates $C_i$ in the Heisenberg picture. 
For convenience, we label the gates in the order in which they are applied to the operator. Note that since Clifford gates both include the identity and are subsets of general unitary gates, the above expression can represent any quantum circuit with $N$ gates.

Without loss of generality, through circuit compilation we can consider circuits (\ref{eq:exp_val_m}) containing only Pauli rotations $U_i \equiv U_i(\theta_i) = \exp(-i \theta_i P_i / 2) $, where $P_i$ are Pauli operators, and, further, that the angles satisfy $|\theta_i| \leq \pi/4$.  
The latter is because for any $|\theta_i|>\pi/4$, we can use the angle transformation $U_i(\theta_i) = U_i(\tilde{\theta_i} + k \pi /2) C_{U_i}$, where $C_{U_i} = \exp(-i k \pi P_i / 4)$ ($k \in \mathbb{Z}$) is again a Clifford gate.
 Similarly, because Clifford gates map the Pauli group onto itself \cite{Gottesman:1998}, every gate $C_i$ can be applied to the observable $O$ and all $P_{j<i}$ without increasing the number of terms, leaving only Pauli rotations with modified $P_i$. 
Applying these two transformations is efficient and yields the Clifford interaction picture, whereby Eq.~\eqref{eq:exp_val_m} becomes
\begin{align}
      \langle O \rangle = \langle 0^{\otimes n}|   {U}_N(\theta_N)^{\dag} \cdots  {U}_1(\theta_1)^{\dag} O {U}_1(\theta_1)  \cdots {U}_N(\theta_N)  |0^{\otimes n} \rangle \label{eq:clifford_interaction}
 \end{align}
with all $|\theta_i| \leq \pi/4$, and the $U_i$ have been transformed by the Cliffords. 


To reduce the cost of evaluating Eq.~(\ref{eq:clifford_interaction}), we first note that many circuits possess a certain structure that reduces the number of gates that affect the final expectation value. 
Specifically, the application of a Pauli rotation to a Pauli operator yields:
\begin{equation}
e^{i \theta P / 2} O e^{-i \theta P / 2} = 
\begin{cases}
O, & [P, O] = 0, \\
\cos(\theta) O + i \sin(\theta) P O & \{P, O\} = 0.
\end{cases} \label{eq:evolution}
\end{equation}
Therefore, only gates that anticommute with the observable increase the number of Pauli terms needed to represent the Heisenberg-evolved observable and thus contribute to the computational cost. The full circuit is applied by iterating Eq.~(\ref{eq:evolution}), that is, at step $i+1$, the evolved observable is $O_{i+1} =U_{i+1}(\theta_{i+1})^{\dag} O_i U_{i+1}(\theta_{i+1}) = \sum_j  U_{i+1}(\theta_{i+1})^{\dag}\sigma_{i,j} U_{i+1}(\theta_{i+1})$, where $O_i = \sum_j \sigma_{i,j}$ and $\sigma_{i,j}$ are Pauli operators. The worst-case scaling of this method is $2^N$, which is attained only if all $P_i$ anticommute with $O$ and commute with each other. For random circuits, we can expect that $P_{i+1}$ will commute on average with half of the Pauli terms in the evolved observable and generate $M_{i+1} = 3/2 M_i$ Pauli terms, leading to an average scaling of $(3/2)^n$.

In this work, we do not aim to propose an exact or rigorously $\delta$-approximate method with an asymptotic exponential scaling better than that already available in the literature. Rather, we are interested in computing expectation values approximately, but in a systematically improvable way, via a perturbation expansion.
For this purpose, we note that the ratio between coefficients of the $O$ and $P O$ branches in Eq.~\eqref{eq:evolution} is $\tan(\theta)$. Since $|\theta| \leq \pi/4$, we know that $|\tan(\theta)| \leq 1$, i.e., the two branches will be weighted equally in the worst case, but otherwise, the term involving the unmodified Pauli operator $O$ will be larger. This motivates an ordering of terms by the number of multiplications by $\sin(\theta_i)$, which we call a perturbation order $k$. We can rewrite the mean value (\ref{eq:exp_val_m}) as
\begin{align}
\langle O \rangle &= \sum_{k=0}^{\tilde{N}} E^{(k)}, \\
E^{(k)} &= i^k \sum_{1 \leq j_1 < j_2 < \cdots < j_k \leq \tilde{N}} c_{j} \sin(\theta_{j_1}) \cdots \sin(\theta_{j_k}) \langle 0^{\otimes n}| P_{j_k} \cdots P_{j_1}O |0^{\otimes n} \rangle,\label{eq:pert_k}
\end{align}
where $\tilde{N} \leq N$ is the highest order of perturbation and $c_j$ are products of cosine functions,
\begin{equation}
c_j = \prod_{l\notin \{ j_1, \dots , j_k\}}^N \cos^{\xi_l}(\theta_l). \label{eq:c_j}
\end{equation}
In Eq.~(\ref{eq:c_j}), $\xi_l = 0$ if the Pauli operator $P_l$, associated with the Pauli rotation gate $U_l(\theta_l)$, commutes with $P_{j_m} \cdots P_{j_1} O$ for $j_m<l<j_{m+1}$ and $\xi_l=1$ otherwise. The goal then is to explore the truncated series
\begin{equation}
\langle O \rangle^{(K)} = \sum_{k=0}^{K} E^{(k)}, \quad K \leq \tilde{N}
\end{equation}
as an approximation for $\langle O \rangle$. This series expansion resembles the series expansion of time-dependent perturbation theory, thus we refer to this method as Clifford perturbation theory. The angles $|\tan(\theta)|$ serve as a measure of the non-Clifford nature of the gates and are analogous to perturbation parameters, and the truncated series is thus most accurate for near-Clifford circuits where $|\tan(\theta)| \sim 0$. Note that because the circuit is finite and discrete, unlike in perturbation theories associated with continuous Hamiltonian evolution, the series always truncates at $K=\tilde{N}$ and cannot diverge.

For simplicity, let us consider an example in which all Pauli operators $P_i$ anticommute with $O$ and commute with each other. In this case, $\tilde{N} = N$ and
\begin{equation}
E^{(k)} = i^{k} \left( \prod_{i=1}^{N} \cos(\theta_i)\right) \sum_{1 \leq j_1 < j_2 < \cdots < j_k \leq N} \tan(\theta_{j_1}) \cdots \tan(\theta_{j_k}) \langle 0^{\otimes n}| P_{j_k} \cdots P_{j_1}O |0^{\otimes n} \rangle.\label{eq:pert_k_special}
\end{equation}
On one hand, we can rely on the fact that each term at order $k$ will have an additional factor of $\tan(\theta_{j_k})$ compared to a term at order $k-1$. On the other hand, there are $M_k = \binom{N}{k}$ terms at order $k$. In practice, however, many terms in Eq.~(\ref{eq:pert_k_special}) might be zero regardless of the value of $\theta$. For example, for a random Pauli operator $P$ with a weight $w$ the probability of $\langle 0^{\otimes n}| P |0^{\otimes n}\rangle \neq 0$ is $3^{-w}$. Since this possibility is not known before the Pauli operator $P_{j_k} \cdots P_{j_1}O$ in (\ref{eq:pert_k_special}) is computed, we choose to study this perturbative expansion numerically. In Appendix A, we present a more detailed analysis of the perturbative treatment as applied to random quantum circuits.

We note that a work closely related to ours recently appeared, where the authors consider the truncated Fourier series as an approximation to the quantum mean value problem \cite{Nemkov_Fedorov:2023}. However, their expansion orders terms in Fourier levels that are defined differently from the perturbation series considered in this work. In addition, another perturbative method \cite{Mitarai_Fujii:2022} was proposed for approximately optimizing VQE parameters, in this case, the angles of Pauli rotation gates, in a circuit composed of alternating Pauli and Clifford gates. There, the authors expanded the mean value to second order in the parameters, thereby effectively reducing the problem to a classically efficient Clifford-circuit simulation of the expectation value, its gradient, and Hessian with respect to the parameters.

\section{Numerical examples\label{sec:res}}

The Clifford perturbation method was implemented using Qiskit \cite{Qiskit} and studied on two different problems.
First, we applied the method to evaluate the QAOA cost function of the combinatorial Max E3LIN2 problem \cite{Farhi_Gutmann:2015,Bravyi_Howard:2019}. Second, we simulated mean values of a Clifford circuit that was subject to coherent noise, a common source of errors in modern-day implementations of quantum circuits \cite{Ouyang:2021}.

\subsection{QAOA applied to the Max E3LIN2 problem}
The Max E3LIN2 problem is a combinatorial optimization task of finding bitstrings $z_1, z_2, ..., z_n$ that maximize the cost function
\begin{equation}
C_z = \frac{1}{2}\sum_{1 \leq u < v < w \leq n} d_{uvw} z_u z_v z_w, \label{eq:qaoa_cost}
\end{equation}
determined by coefficients $d_{uvw} \in \{0, \pm 1\}$ and a parameter $D$ that defines the number of occurrences of 
a given $z_i$ in the sum (\ref{eq:qaoa_cost}). QAOA solves this task by constructing an ansatz
\begin{equation}
|\mathbf{\beta}, \mathbf{\gamma}\rangle = \prod_{j=1}^{p} e^{-\beta_j \sum_{i}^{n} X_i} e^{-i \gamma_j C} H^{\otimes n} |0^{\otimes n} \rangle, 
\end{equation}
and optimizing the expectation value $\langle C \rangle = \langle \mathbf{\beta}, \mathbf{\gamma}| C |\mathbf{\beta}, \mathbf{\gamma}\rangle$, where $C$ is obtained from $C_z$ by replacing the bits $z_u$ in Eq.~(\ref{eq:qaoa_cost}) by Pauli operators $Z_u$. In what follows, we evaluate the expectation value
\begin{equation}
\langle C \rangle = \frac{1}{2}\sum_{1 \leq u < v < w \leq n} d_{uvw} \langle \mathbf{\beta}, \mathbf{\gamma}| Z_u Z_v Z_w |\mathbf{\beta}, \mathbf{\gamma}\rangle \label{eq:qaoa_cost_split}
\end{equation}
by applying the algorithm introduced earlier to individual terms on the right-hand side of (\ref{eq:qaoa_cost_split}).

\begin{figure}[!pth]
\includegraphics[width=0.5\textwidth]{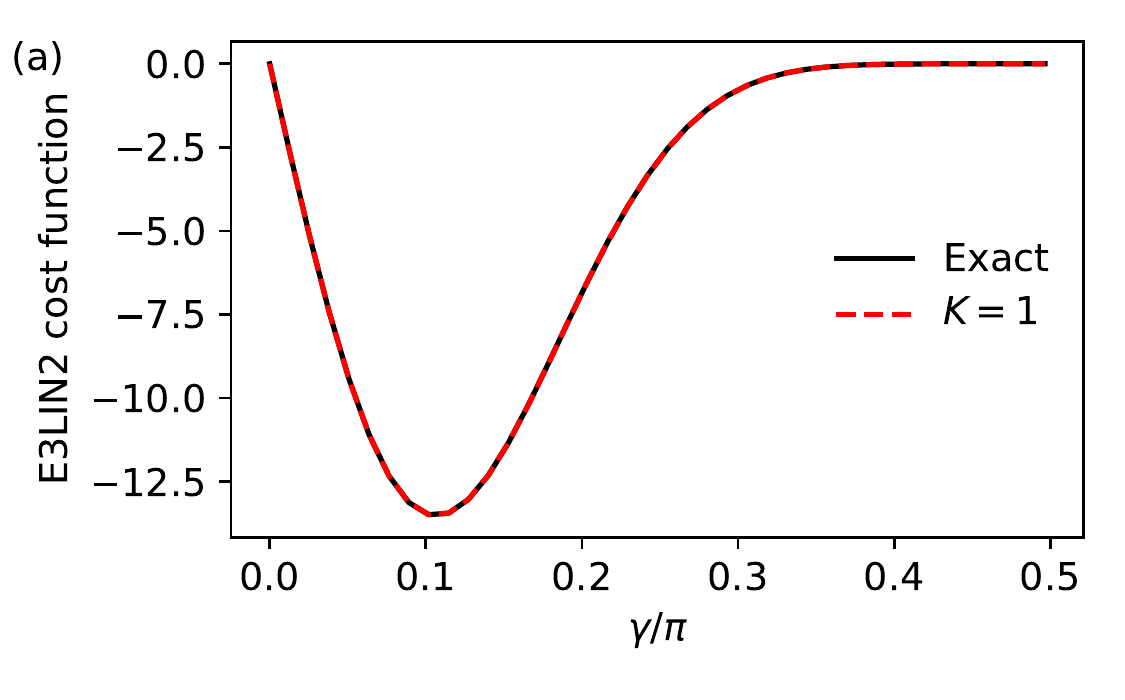}
\includegraphics[width=0.5\textwidth]{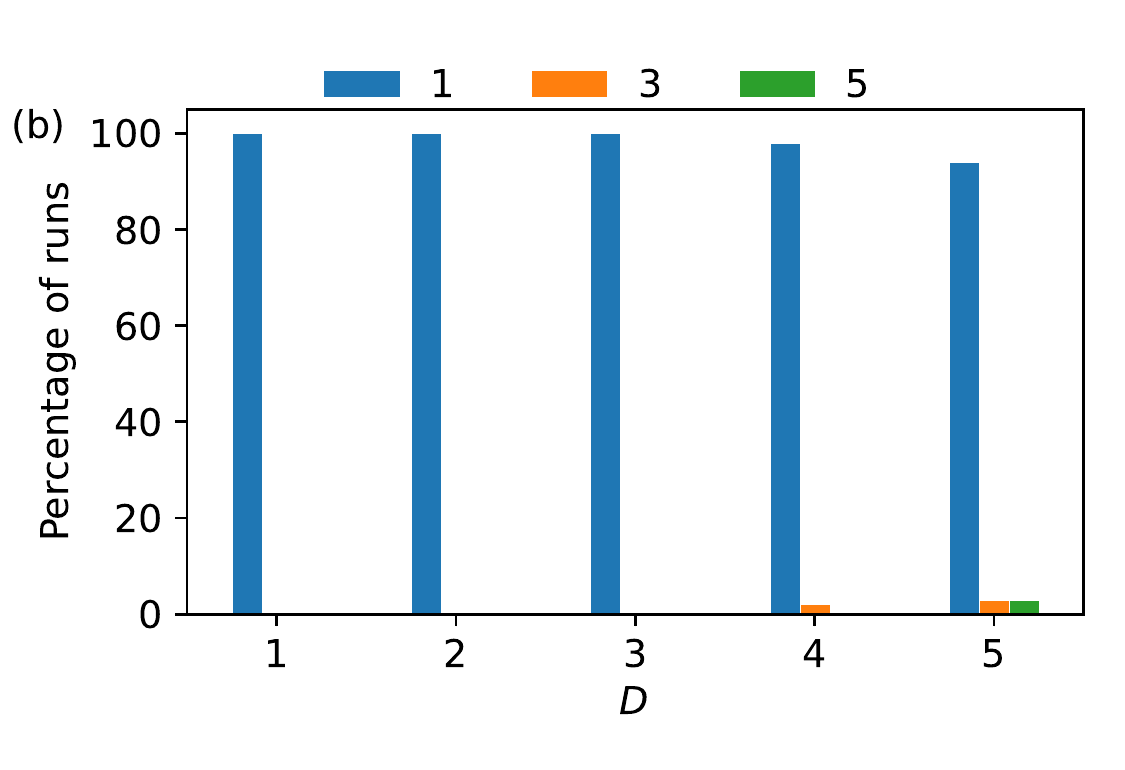}
\caption{\label{fig:QAOAExample}(a) QAOA cost function (\ref{eq:qaoa_cost_split}) for the E3LIN2 problem with $n=50$ qubits and $D=4$. Only positive values of $\gamma$ are shown because of the symmetry relation $C(-\gamma) = -C(\gamma)$. (b) Histogram over 100 runs of the maximum order $K$ with a nonzero contribution to the cost function (\ref{eq:qaoa_cost_split}) with different values of $D$ and $n=50$ qubits.}
\end{figure}

Following Bravyi et al. \cite{Bravyi_Howard:2019}, we consider an example with $n=50$ qubits, $p=1$ (one-layer ansatz), and $\beta = \pi / 4$.
The expectation value $\langle C \rangle$ is shown as a function of $\gamma$ in Fig.~\ref{fig:QAOAExample} (top), where each point took only a few seconds to compute on a laptop. 
In comparison, in \cite{Bravyi_Howard:2019}, generating the same data as in Fig.~\ref{fig:QAOAExample}a with the low-rank stabilizer method was reported to take about 3 days. This reflects the fact that whereas sampling from the probability distribution $p(x) = |\langle x | \beta, \gamma\rangle|^2$ is thought to be generically hard, the expectation values (\ref{eq:qaoa_cost_split}) can be computed in polynomial time due to the constant
depth nature of the circuit, which means that there is a constant size light-cone (with system size). 
Therefore, this model illustrates the significant simplification afforded if one targets the computation of expectation values with a method that takes advantage of the light-cone structure and the shallow depth. 
Indeed, even though the whole circuit is composed of $N = \lfloor n \times D / 3 \rfloor = 66$ non-Clifford gates, only a fraction of those, namely $\tilde{N} \leq 3(D-1)+1$, anticommute with the individual Pauli terms in the observable \cite{Rall_Kretschmer:2019}.


However, in Fig.~\ref{fig:QAOAExample}a we also show that in this case there is no error if we truncate the perturbation order to $K=1$, regardless of the value of $\gamma$. This arises from the algebraic structure of the gates and observable, which mean that not all operators in the observable lightcone yield a non-zero expectation value, giving additional savings. Prompted by this result, we evaluated randomly generated circuits for $1 \leq D \leq 5$ and for each instance evaluated the number of instances with contributions from higher orders to the expectation value. As seen in Fig.~\ref{fig:QAOAExample}b, almost all instances require only the first-order ($K=1$) perturbation contribution, whereas at higher $D$ there are examples with nonzero third- and fifth-order contributions. These contributions will additionally be weighted by factors of $\tan(\theta)$ which will usually make such terms small. Thus in certain shallow QAOA circuits~\cite{Zhuang_Liu:2021}, the perturbation expansion can take advantage of additional truncations and approximations to the observable lightcone, speeding up the classical simulation.

\subsection{Coherent error in Clifford circuits}
Clifford gates are essential for the implementation of quantum error correction \cite{Bravyi_Maslov:2022} and for the validation of new quantum hardware \cite{Cross_Gambetta:2019}. Quantum coherent errors introduce a bias in the rotation angles, which implies that simulating such noisy circuits can become difficult because the gates are no longer Clifford. Here, we show that our method can be of use in such cases, especially in the limit of small errors in the rotation angles. Moreover, we show in Appendix B that our approach is applicable to incoherent errors as well, including those represented by Pauli and amplitude- or phase-damping channels. 

We construct a Clifford circuit
\begin{equation}
U = \prod_{j=1}^{p} B_j A_j \label{eq:CliffordLayers}
\end{equation}
from alternating layers of one-qubit
\begin{equation}
A = e^{- \frac{i}{2} \sum_{i=1}^n\theta_{i} \sigma_{i}} \label{eq:Clifford_1q}
\end{equation}
and two-qubit
\begin{equation}
B = e^{- \frac{i}{2} \sum_{i=1}^{n/2}\theta_{i} P_{i} }, \quad P_{i} = \sigma_{i_1}^{\prime} \sigma_{i_2}^{\prime \prime} \label{eq:Clifford_2q}
\end{equation}
Pauli rotation gates, where $\sigma, \sigma^{\prime}, \sigma^{\prime \prime} \in \{ X, Y, Z\}$ are selected randomly. In (\ref{eq:Clifford_2q}), all $i_k$ in the pairs are distinct and chosen randomly, i.e., exactly $n/2$ random disjoint pairs appear within a single two-qubit layer. The rotation angles are set to $\theta \in \{\pm\pi/2, \pm\pi \}$, so that all gates are Clifford. Then, all rotations are distorted ($\theta \rightarrow \theta + \delta\theta$) by $\delta\theta$, which represents the error. We use $n=50$ qubits and measure the expectation value of $Z_1 Z_{26}$.

\begin{figure}[!pth]
\raggedright
\includegraphics[width=0.48\textwidth]{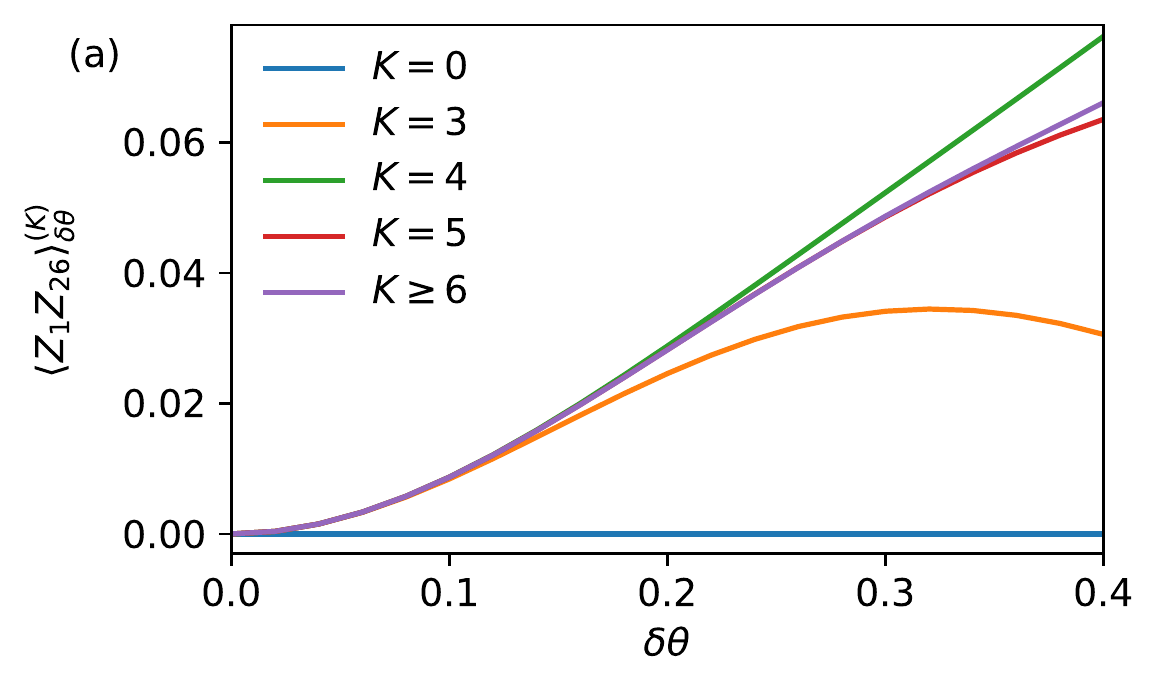}
\includegraphics[width=0.48\textwidth]{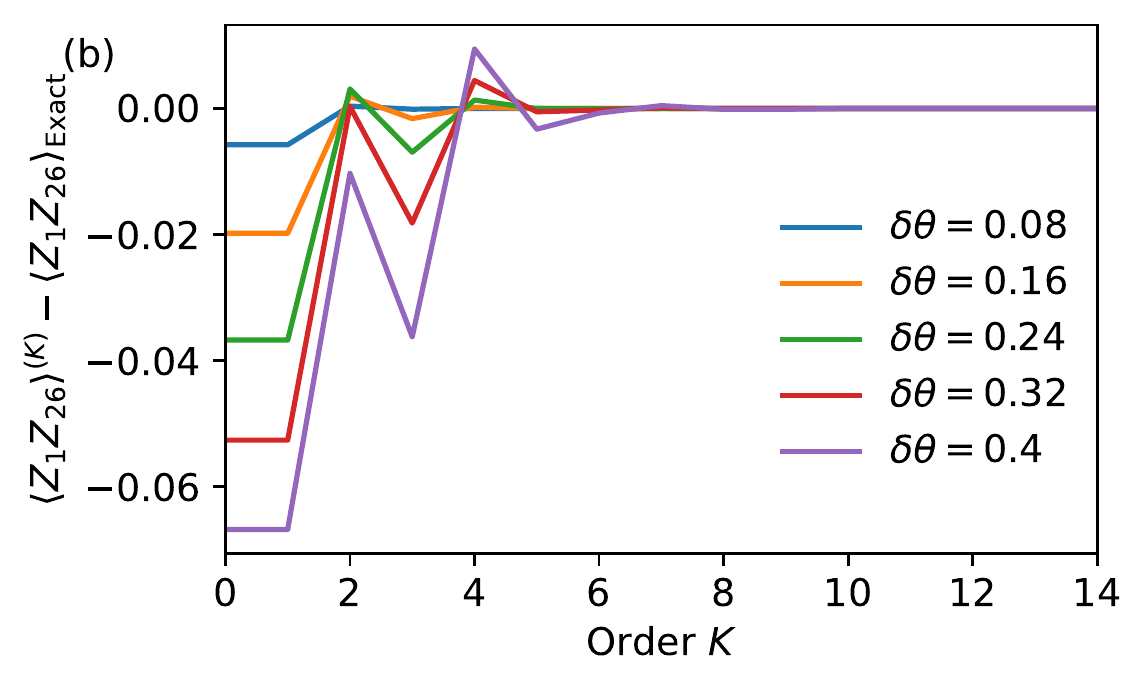}
\includegraphics[width=0.48\textwidth]{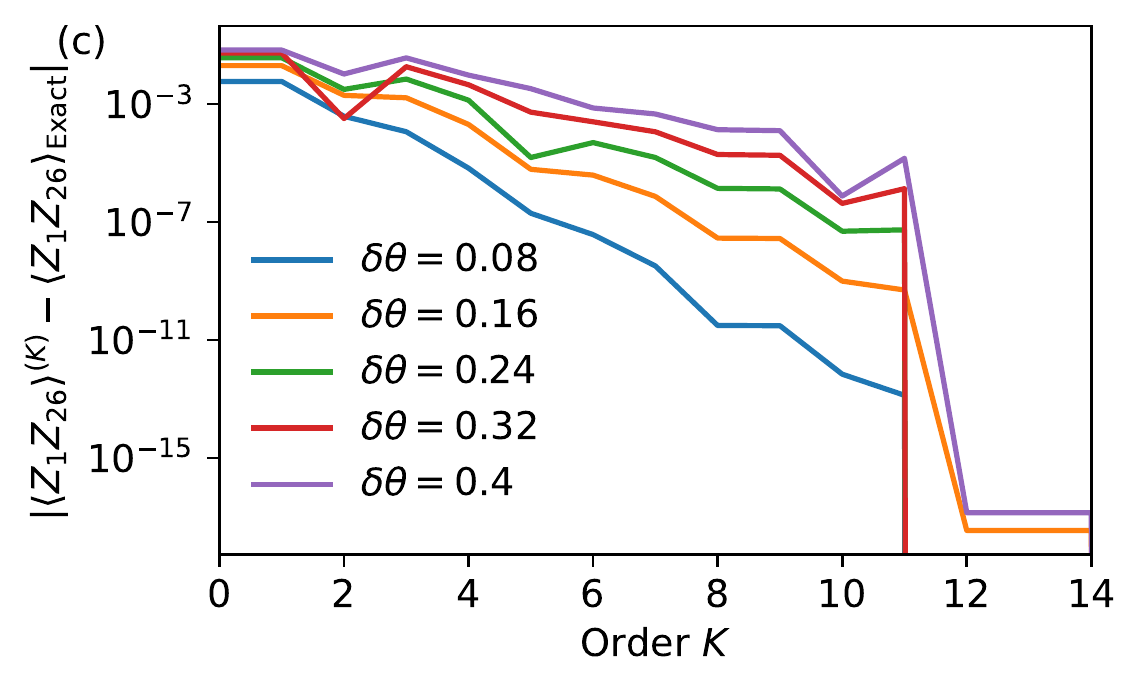}
\includegraphics[width=0.48\textwidth]{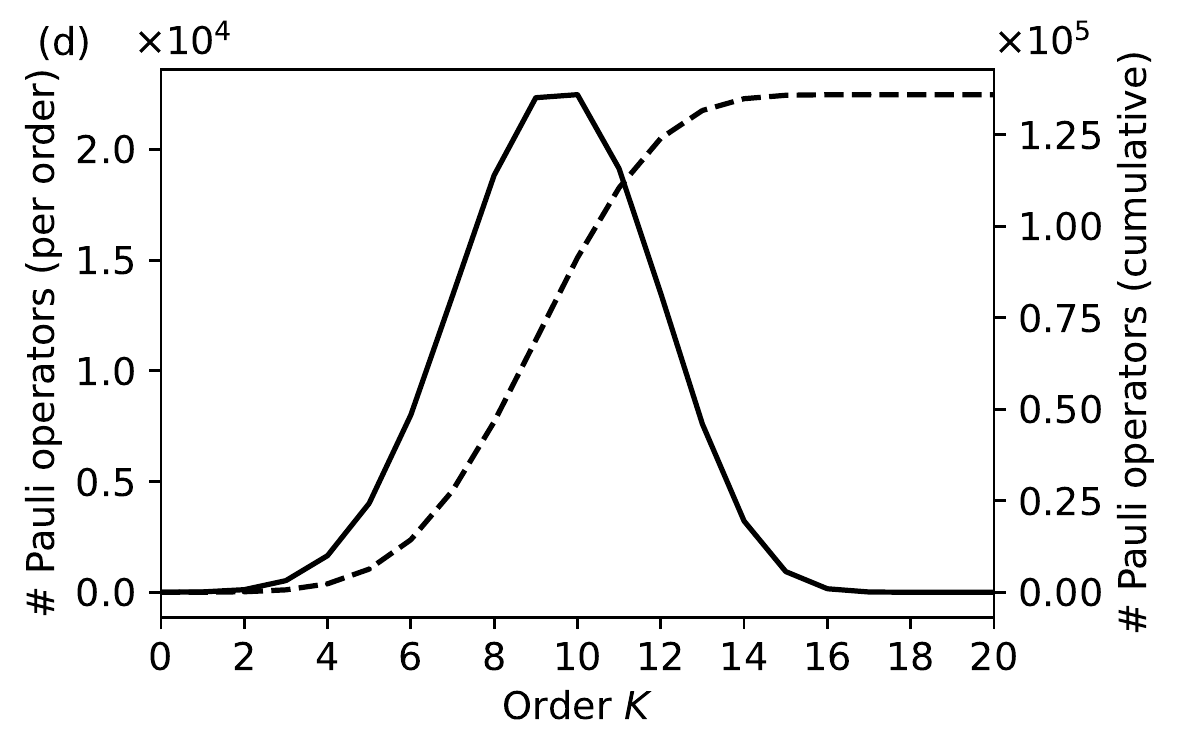}
\caption{\label{fig:CliffordLayers}$\langle Z_1 Z_{26} \rangle$ evaluated for a Clifford circuit defined by Eqs.~(\ref{eq:CliffordLayers})--(\ref{eq:Clifford_2q}) with $n=50$ qubits, $p=4$, and a coherent error defined by $\delta\theta$ (see text). (a) Expectation value as a function of $\delta \theta$ for several perturbation orders $K$; (b, c) error in the expectation value as a function of the perturbation order (on linear (b) and logarithmic (c) scales); (d) cumulative (dashed, right-hand axis) and per-order (solid, left-hand axis) number of Pauli operators generated in the evaluation of $\langle Z_1 Z_{26} \rangle$.}
\end{figure}

Figure~\ref{fig:CliffordLayers}a plots the expectation value as a function of $\delta\theta$ for different values of perturbation order $K$. As $\delta\theta$ increases, the expectation value under coherent noise diverges from its noiseless result. Unsurprisingly, the low-order perturbation approach works better for small $\delta \theta$, whereas for larger $\delta\theta$ higher-order perturbation terms are needed. This is shown clearly in Figs.~\ref{fig:CliffordLayers}b, c. In this example with $p=4$ ($300$ gates), we can simulate the expectation values accurately, e.g., with error $<10^{-2}$, already with $K=4$ or even $K=2$ for the smallest values of $\delta\theta$. For comparison, the total number of Pauli operators generated by evolving the observable as $U^{\dag} Z_1 Z_{26} U$ is $135930$, whereas only $2313$ operators are generated for orders up to $K=4$ (see Fig.~\ref{fig:CliffordLayers}d).


\begin{figure}[!pth]
\centering
\includegraphics[width=0.58\textwidth]{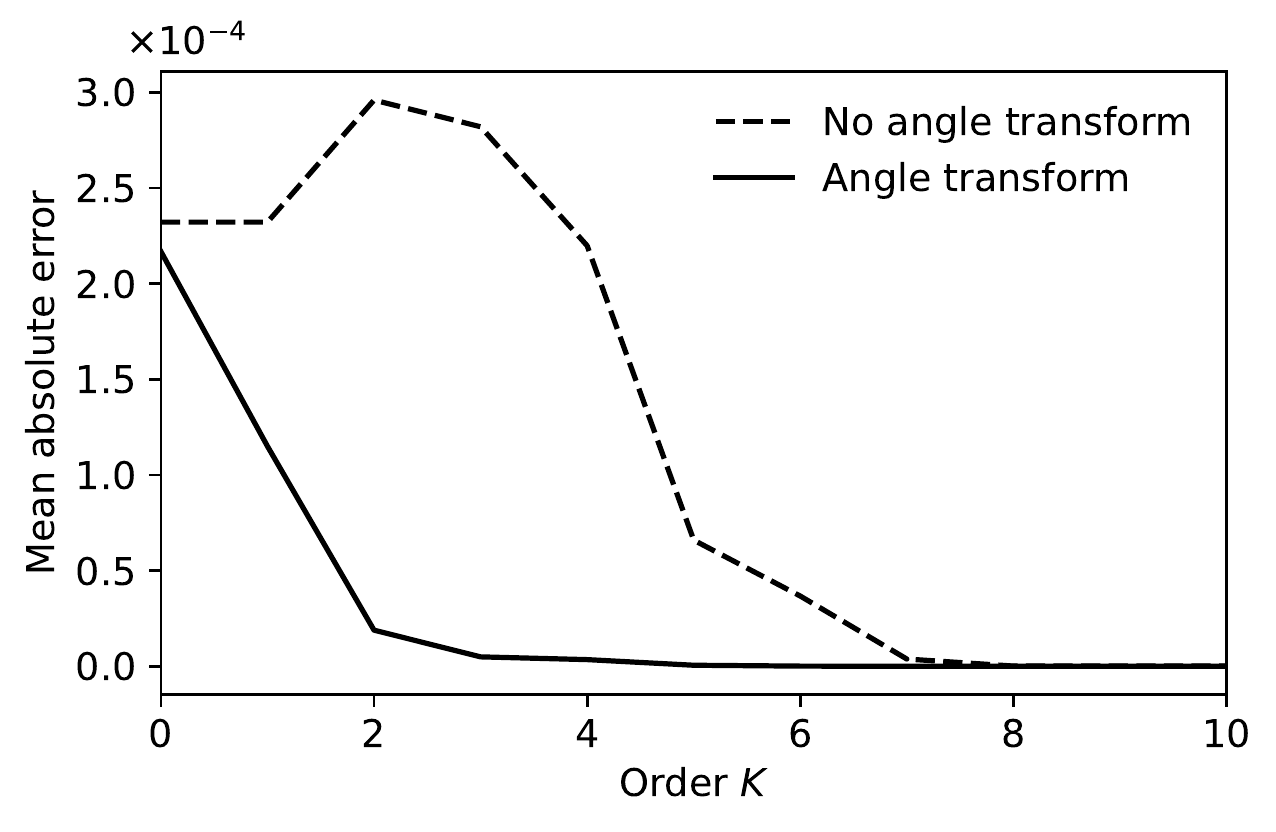}
\caption{\label{fig:InteractionPicture}Mean absolute values of errors in $\langle Z_1 Z_{26} \rangle$ for 1000 random circuits generated according to Eqs.~(\ref{eq:CliffordLayers})--(\ref{eq:Clifford_2q}) with $n=50$ qubits, $p=3$, and $\delta\theta=0.2$. The results are shown for two different calculations: one in which the perturbation theory is applied directly to the Pauli rotation gates (``no angle transform'') and the other in which the rotation gates are first transformed so that $|\theta|<\pi/4$ (``angle transform''). }
\end{figure}

Next, we illustrate the advantage of carrying out perturbation theory after using the angle transformation on the circuit so that $|\theta| \leq \pi/4$, as outlined in Sec.~\ref{sec:method}.
Figure~\ref{fig:InteractionPicture} compares the result from Eq.~\ref{eq:clifford_interaction} with and without the angle transformation. Here, the mean absolute error is computed from 1000 examples constructed with $n=50$ qubits and $p=3$ layers. Low perturbation orders contribute little to the final expectation values if the rotation gates are not transformed, whereas an opposite trend is observed within the angle transformed picture.

Finally, let us consider an example where performing the exact computation would require extensive computational resources, beyond what can be done with a basic implementation on a personal computer. We set $n=100$, $p=3$ (450 gates), and $\delta\theta=0.2$. The calculation up to perturbation order $K=7$ took approximately 2 hours on a laptop and generated around $10^7$ Pauli terms. Given that the expectation values for all $K \geq 4$ are similar, we can estimate $\langle Z_1 Z_{50} \rangle \approx \langle Z_1 Z_{50} \rangle^{(7)} = 0.0023$. 

\begin{figure}[!pth]
\centering
\includegraphics[width=0.63\textwidth]{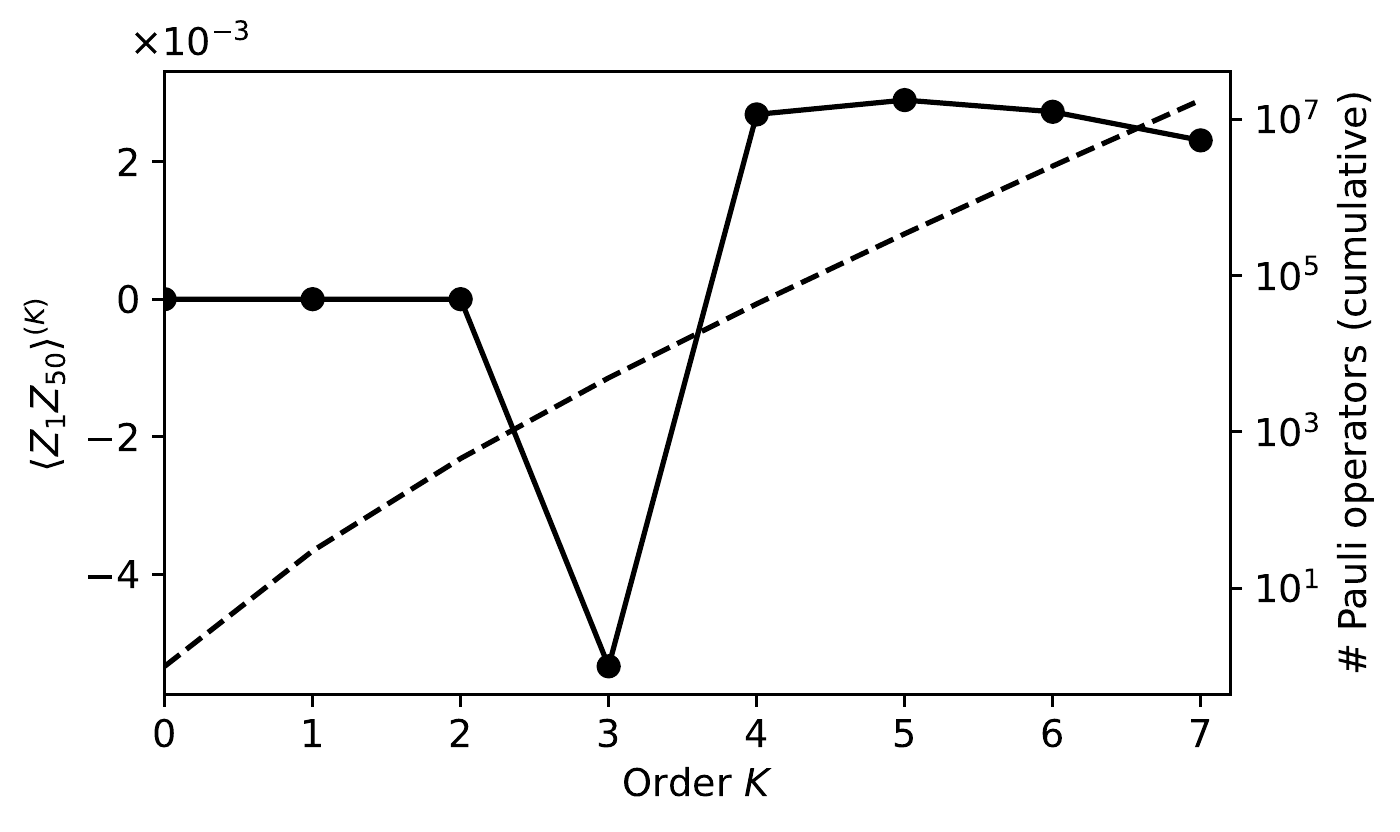}
\caption{\label{fig:LargeCliffordCircuit}$\langle Z_1 Z_{50} \rangle$ expectation value (solid, left-hand $y$-axis) for a Clifford circuit ($n=100$, $p=3$) with coherent error ($\delta\theta = 0.2$) computed up to perturbation order $K=7$. The dashed line (right-hand $y$-axis) shows the total number of Pauli operators generated up to order $K$.}
\end{figure}

\section{Conclusion\label{sec:conclusion}}

To conclude, we have introduced a Clifford-based perturbation theory that enables the efficient computation of expectation values of near-Clifford quantum circuits. We studied the merits of this approximate scheme in the context of QAOA and Clifford circuits subject to coherent noise. These numerical examples helped illustrate the two sources of speed-up in our method: First, the fact that in some circuits, for example shallow QAOA, only low-order terms in the perturbation expansion are found to be non-zero, and second, the fact that sufficiently small angles of Pauli rotation gates suppress contributions from higher-order terms. In conjunction, these simplifications enable the practical determination of expectation values of observables of large near-Clifford circuits, for example, with over a hundred qubits and hundreds of gates, using only laptop computational resources.

\section{Acknowledgments}

We acknowledge informative discussions with Steve Flammia, Ali Lavasani, David Gosset, Sergey Bravyi, and Alex Dalzell. TB and GKC were supported by the US Department of Energy, Office of Science, Office of Advanced Scientific Computing Research and Office of Basic Energy Sciences,
Scientific Discovery through Advanced Computing (SciDAC) program under Award Number DE-SC0022088. TB acknowledges financial support from the Swiss National Science Foundation through the Postdoc Mobility Fellowship (grant number P500PN-214214). GKC is a Simons Investigator in Physics.

\appendix
\section{Random quantum circuits}
\begin{figure}[!pth]
\centering
\includegraphics[width=0.63\textwidth]{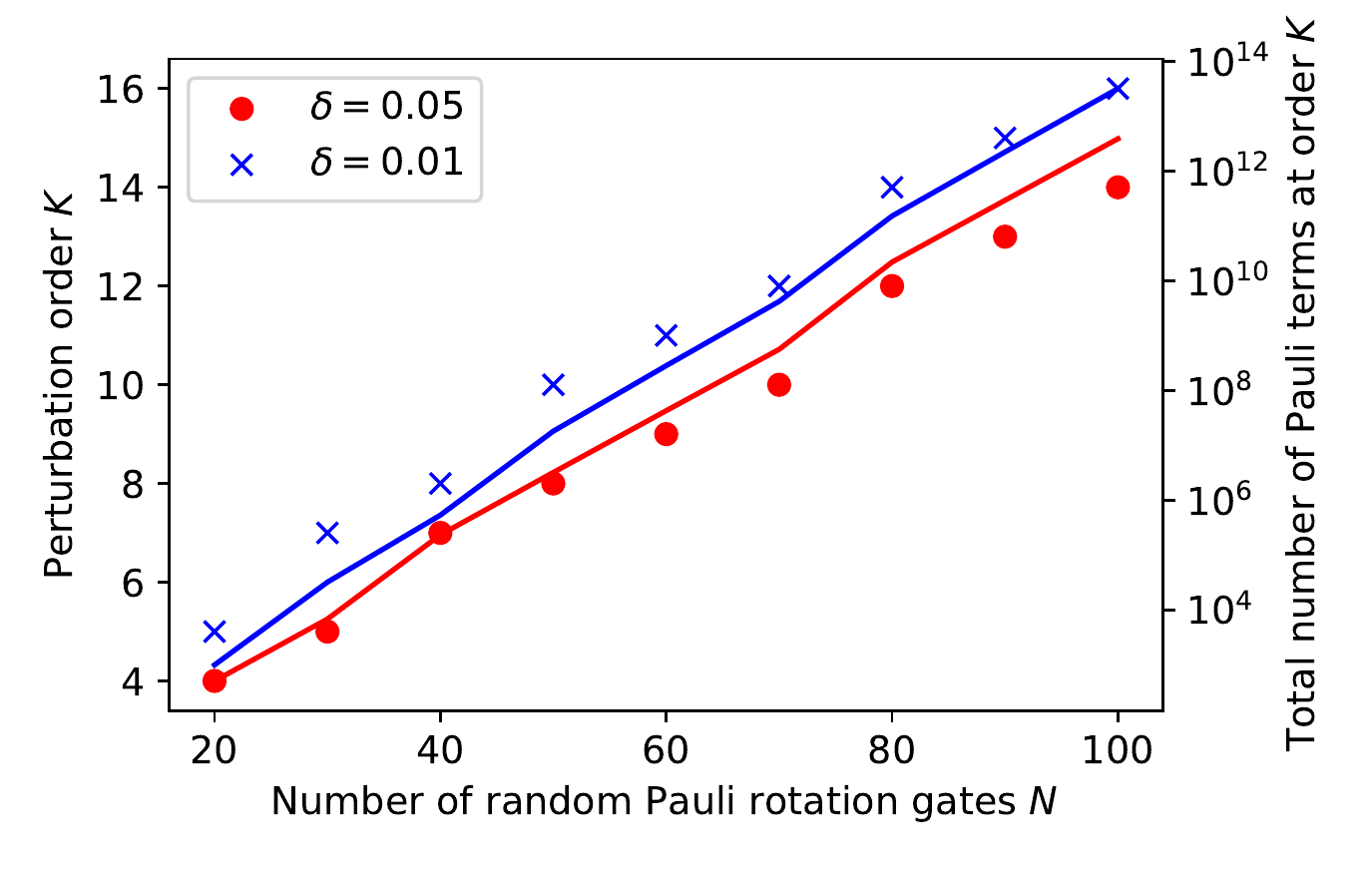}
\caption{\label{fig:RandomCircuitOrder}Minimal value of the perturbation order $K$ that satisfies Eq.~(\ref{eq:bound}) for different numbers of gates $N$, angle $\theta = 0.2$, and $\delta=0.05$ (red markers) or $\delta=0.01$ (blue markers). Solid lines (right-hand $y$-axis) correspond to $\sum_{k=0}^{K}M^{(k)}$ for each pair of $N$ and $K$.}
\end{figure}

Let us consider a random quantum circuit composed of Pauli rotations with a fixed small angle $\theta_i = \theta$, such that $\cos(\theta) \approx 1$. Then,
\begin{equation}
E^{(k)} = i^k \sin^{k}(\theta) \sum_{1 \leq j_1 < \cdots <j_k \leq \tilde{N}} \langle 0^{\otimes n}| P_{j_k} \cdots P_{j_1}O |0^{\otimes n} \rangle.
\end{equation}
In the remainder, we will not discuss the value of $i^k \langle 0^{\otimes n}| P_{j_k} \cdots P_{j_1}O |0^{\otimes n} \rangle \in \{0, \pm 1\}$, but rather focus on the average value of the sum of absolute values of the coefficients
\begin{equation}
|E|^{(k)} = M^{(k)} |\sin(\theta)|^k,
\end{equation}
where $M^{(k)}$ denotes the average number of Pauli operators at perturbation order $k$. We can deduce $M^{(k)}$ by assuming that the Pauli rotation gate $U_{i+1}(\theta)$ commutes with each term in the evolved observable $O_{i}$ with a probability of 0.5. Let us label one Pauli operator in $O_{i}$ as $o_{i}^{(k)}$, where $k$ indicates the number of times it has been multiplied by $\sin(\theta)$, i.e., its order. If $[P_{i+1}, o_i^{(k)}] = 0$, according to Eq.~(\ref{eq:evolution}), $U_{i+1}(\theta)^{\dag} o_i^{(k)} U_{i+1}(\theta) = o_i^{(k)}$, i.e., the number of terms remains the same. Otherwise, if $\{P_{i+1}, o_i^{(k)}\} = 0$, $U_{i+1}(\theta)^{\dag} o_i^{(k)} U_{i+1}(\theta) = \cos(\theta) o_i^{(k)} + i \sin(\theta) P_{i+1} o_i^{(k)}$. Here, the number of Pauli operators of order $k$ remains the same, but the number of terms of order $k+1$ is increased by 1. This results in the following recursive formula
\begin{equation}
M_{i}^{(k)} = M_{i-1}^{(k)} + \frac{1}{2}  M_{i-1}^{(k-1)}, \label{eq:Mk_recursive}
\end{equation}
where $M_{i}^{(k)}$ denotes the number of terms of order $k$ in the observable $O_{i}$ evolved up to step $i$ and $M_0^{(k)} = \delta_{k0}$. By induction, we can prove the following closed-form expression
\begin{equation}
M^{(k)} = M_N^{(k)} = 2^{-k} \binom{N}{k}, \label{eq:Mk_closed}
\end{equation}
which agrees with the average total number of Pauli operators $M = \sum_{k=0}^{N} M^{(k)} = (3/2)^N$.
Then,
\begin{equation}
|E|^{(k)} =|\sin(\theta) / 2|^k \binom{N}{k}
\end{equation}
and
\begin{equation}
|O|^{(K)} = \sum_{k=0}^{K} |E|^{(k)} = \sum_{k=0}^{K} |\sin(\theta) / 2|^k \binom{N}{k}.
\end{equation}
In particular, 
\begin{equation}
|O| = |O|^{(N)} = (1 + |\sin(\theta) / 2|)^N.
\end{equation}
Let us now consider an upper bound on the relative error
\begin{equation}
 \frac{|O| - |O|^{(K)}}{|O|} < \delta. \label{eq:bound}
\end{equation}
Specifically, we wish to find the smallest value of perturbation order $K$ that satisfies the equation above for given $N$, $\theta$, and $\delta$. As an example, Fig.~\ref{fig:RandomCircuitOrder} shows $K$ as a function of $N$ for $\delta = 0.01, 0.05$, and $\theta = 0.2$.

\section{A note on incoherent errors}
Several basic models of errors \cite{Nielsen_Chuang:2010} can be represented as a Pauli channel
\begin{equation}
\tilde{\rho} = (1 - s_x -  s_y - s_z) \rho + s_x X \rho X + s_y Y \rho Y + s_z Z \rho Z, \label{eq:PauliChannel}
\end{equation}
including bit flip errors
\begin{equation}
\tilde{\rho} = (1-s) \rho + s X \rho X,
\end{equation}
phase flip errors
\begin{equation}
\tilde{\rho} = (1-s) \rho + s Z \rho Z,
\end{equation}
and a depolarizing channel
\begin{equation}
\tilde{\rho} = (1 - s) \rho + \frac{s}{3} (X \rho X + Y \rho Y + Z \rho Z).
\end{equation}
Pauli channels can be implemented with our method deterministically and at almost no additional computational cost. First, we note that the Pauli operators and the scaling factor in Eq.~(\ref{eq:PauliChannel}) can be applied directly to the observable (i.e. in the Heisenberg picture) instead of on the density operator. Second, the conjugation of one Pauli operator by another results only in a potential sign change:
\begin{equation}
\sigma_{\text{Error}} \sigma_{\text{Observable}} \sigma_{\text{Error}} = \pm \sigma_{\text{Observable}},
\end{equation}
where the sign depends on whether the two Pauli operators commute. Therefore, a one-qubit Pauli channel applied to a Pauli operator will only introduce a scaling factor 
\begin{equation}
\eta = 1 - s_x -  s_y - s_z \pm s_x  \pm s_y  \pm s_z.
\end{equation}

Another example is the amplitude-damping error \cite{Nielsen_Chuang:2010}, which can be represented as:
\begin{equation}
\tilde{\rho} = E_0^{\dagger} \rho E_0 + E_1^{\dagger} \rho E_1, \label{eq:AmpDampingChannel}
\end{equation}
where
\begin{align}
E_0 &= \begin{pmatrix} 1 & 0 \\ 0 & \sqrt{1-\lambda} \end{pmatrix}, \\
E_1 &= \begin{pmatrix} 0 & \sqrt{\lambda} \\ 0 & 0 \end{pmatrix}.
\end{align}
Equation~(\ref{eq:AmpDampingChannel}) applied to single-qubit Pauli operators yields:
\begin{align}
\tilde{I} &= I \label{eq:I_amp_damping}\\
\tilde{X} &= \sqrt{1-\lambda}X  \label{eq:X_amp_damping}\\
\tilde{Y} &= \sqrt{1-\lambda}Y  \label{eq:Y_amp_damping}\\
\tilde{Z} &= (1-\lambda)Z + \lambda I. \label{eq:Z_amp_damping}
\end{align}
$I$, $X$, and $Y$ can be efficiently simulated, but any occurrence of $Z$ will produce twice as many Pauli terms. Fortunately, the two branches in the last equation have different weights if $\lambda$ is small, so the problem can be again treated perturbatively by keeping track of powers of $\lambda$.

Finally, the phase-damping channel can be simulated efficiently. It is similar to the amplitude damping channel [Eq.~(\ref{eq:AmpDampingChannel})] but with
\begin{align}
E_0 &= \begin{pmatrix} 1 & 0 \\ 0 & \sqrt{1-\lambda} \end{pmatrix}, \\
E_1 &= \begin{pmatrix} 0 & 0 \\ 0 & \sqrt{\lambda} \end{pmatrix}.
\end{align}
The application to $Z$ now leaves it unchanged ($\tilde{Z} = Z$), while the action on all other operators is the same as in the amplitude-damping channel [Eqs.~(\ref{eq:I_amp_damping})--(\ref{eq:Y_amp_damping})]. Therefore, the phase-damping channel can be simulated at no additional cost.

\bibliographystyle{quantum}
\bibliography{bibliography}

\end{document}